\documentclass{article}

\usepackage{amssymb,amsfonts,amsmath}
\usepackage{cite,enumerate,float}
\usepackage{color}
\usepackage{tikz}
\usetikzlibrary{arrows,snakes,backgrounds}

\def\be{\begin{eqnarray}}
\def\ee{\end{eqnarray}}
\def\nn{\nonumber}

\def\Tr{{\rm Tr}\,}

\def\cZ{{\cal Z}}

\definecolor{red}{rgb}{1,0,0}
\definecolor{orange}{rgb}{1,0.5,0}
\definecolor{violet}{rgb}{0.7,0,1}

\hoffset= - 1.0in         

\begin{document}

\bigskip

\bigskip

\centerline{\Large{Determinantal representation of colored HOMFLY for double braids
}}

\bigskip

\bigskip

\centerline{\bf Ya. Kononov and A. Morozov$^{1}$}

\bigskip

\centerline{\it $^{1}$ ITEP, Moscow, Russia}

\bigskip

\centerline{ABSTRACT}

\bigskip

{\footnotesize
Starting from the known differential expansions, we express the HOMFLY-PT polynomials of twist and antiparallel double-braid knots in rectangular representations \([r^s]\) as determinants of \(r\times r\) and \(s\times s\) matrices.
We first give an elementary derivation for the figure-eight knot and then re-sum the KNTZ formulas for twist and double-braid evolution.
The resulting matrices are constructed from single-hook evolution coefficients and explicit representation-dependent weights.
These formulas provide a starting point for investigating possible extensions of knot polynomials to KP/Toda \(\tau\)-functions.
Extensions to non-rectangular representations and to more general knot families remain open problems.
}

\bigskip

\bigskip

\section{Introduction}

One of the main problems of quantum field theory (QFT) is an adequate description of
full non-perturbative partition functions \cite{UFN3}.
Since they are given by (functional) integrals with generic measure,
one expects a rich set of Ward identities and, probably,  integrability
and even superintegrability \cite{MMsuper} --
for a clever choice of parameterization in the
infinite space of coupling constants.
An important test of these ideas could be provided by exactly solvable Chern-Simons theory \cite{CS},
where the main subject are non-perturbative Wilson averages, known as colored knot invariants
(Jones \cite{Jones}, HOMFLY-PT \cite{HOMFLYPT}, Kauffman \cite{Kauf} and their exceptional-group counterparts --
as well as Khovanov-Rozansky \cite{KR} and "super"polynomials \cite{superpols}).
However, technical problems still do not allow to do this beyond the simplest family of
torus knots \cite{MMM1} -- which is actually more sensitive to elementary combinatorics
than to a true QFT dynamics.
Far more interesting would be extension to arborescent knots and, to begin with, to
twist and double-braid ones.
In fact, much is understood about them
\cite{IMMM}--\!\!\cite{1906.09971}
-- especially due to the formalism of the differential expansion (DE) \cite{DE}.
Double braids were studied in a special detail, because they provide direct information \cite{1606.06015}
about exclusive  Racah matrices \cite{Racah}.
So far the top summit was the KNTZ formula \cite{KNTZ,1902.04140} for rectangular representations,
describing evolution of HOMFLY along the double-braid family
and attempts of its further generalization to non-rectangular representations \cite{1903.00259}.
Now everything is prepared for the crucial step -- to $\tau$-function description,
and this is what we target at in the present paper.

The true meaning of integrability is the existence of free-field/Gaussian realization
of the non-perturbative (full) partition function \cite{UFN3}.
However, the special and most popularized class of integrable $\tau$-functions involves
realization in terms of free fermions and thus in the form of determinants
and Plucker relations.
What we do now, we demonstrate that KNTZ formula, after additional adjustments,
allows to put rectangularly-colored HOMFLY into determinant form.
This is not yet the full-fledged $\tau$-function with arbitrary time-variables,
but a very important step towards it.

We will follow the simple sequence of knot families

\bigskip

\centerline{
figure-eight knot ${\cal K}=4_1$
$\longrightarrow $  twist knots ${\cal K}={\rm tw}_n$
$\longrightarrow $  double braids ${\cal K}={\rm db}_{m,n}$
}

\bigskip

and for each of them we look at
\underline{symmetric $\longrightarrow$ rectangular} $\longrightarrow$ generic representations.

Essential result concerns the underlined steps, the last one is only speculated about.

\bigskip

{\bf Notation:}

\bigskip

$\{x\}:=x-\frac{1}{x}$, quantum numbers are defined symmetrically $[n]:=\frac{\{q^n\}}{\{q\}}$,
quantum factorial $[n]!:=\prod_{i=1}^n [i]$.
Inverse factorials $\frac{1}{[n]!}$ are defined to vanish for $n<0$.
This is the quantum analogue of the zeros of $1/\Gamma(n+1)$.
All empty products below are equal to one. We use reduced polynomials,
$H_R(\text{unknot})=1$, with independent generic variables $A,q$.

The entries of differential expansions (DE) contain ``differentials'' $D_i:=\{Aq^{i}\}$.
Useful for rectangular representations are peculiar bilinears $Z_{r^s}^{(i)}:=D_{r+i}D_{i-s}$.

Throughout this paper, $R$ denotes the color of the knot polynomial,
whereas $\lambda$ labels one term of its differential expansion.
For example, $\lambda=[2,1]$ occurs in the sum for $R=[2,2]$.
That term is not the polynomial with color $R=[2,1]$.

\section{${\cal K}=4_1$}

We start from the differential expansion (DE), whose explicit weights
are indexed by diagrams $\lambda$ inside a fixed color $R=[r^s]$.
We construct two matrices, $T$ and $Z$, whose paired minors reproduce
these weights. Cauchy--Binet then sums them into a determinant.
Keeping the color matrix $Z$ fixed will then allow us to extend the
construction to twist knots and double braids in
Sections~\ref{sec:twists} and~\ref{sec:double-braids}.

\subsection{Symmetric representations}

For a single-row color $R=[r]$, the figure-eight polynomial is the finite sum
\be
H^{4_1}_{[r]}=1+\sum_{k=1}^r U_k^{[r]},\qquad
U_k^{[r]}:=
\frac{[r]!}{[k]![r-k]!}\prod_{j=0}^{k-1}D_{r+j}D_{j-1}.
\label{eq:symmetric-DE}
\ee
Here $k$ labels the sub-diagram $[k]\subseteq[r]$. For example,
\begin{align*}
H^{4_1}_{[1]}&=1+D_1D_{-1},\\
H^{4_1}_{[2]}&=1+[2]D_2D_{-1}+D_2D_3D_0D_{-1}.
\end{align*}
Any sum $1+\sum_k U_k$ can be written as a determinant:
\be
H^{4_1}_{[r]}
=\det_{1\leq i,j\leq r}\left(\delta_{ij}+U_i^{[r]}\right).
\label{41rdet}
\ee
For $r=2$, this just says
\[
\det\begin{pmatrix}1+U_1&U_1\\U_2&1+U_2\end{pmatrix}
=1+U_1+U_2.
\]
The quadratic terms cancel because the added matrix has rank one.

For a rectangle, the summation diagrams can have several rows.
We need a matrix construction that reproduces their weights as well.
The useful structure is already present in the coefficients of the
differential expansion: each has a product of content factors divided
by the square of a quantum hook product.

\subsection{From the differential expansion to two matrices}
\label{sec:why-two-matrices}

\begin{figure}[h]
\begin{picture}(220,120)(-100,0)

\put(0,0){\vector(1,0){220}}
\put(0,20){\line(1,0){200}}
\put(0,40){\line(1,0){160}}
\put(0,60){\line(1,0){120}}
\put(0,80){\line(1,0){60}}
\put(0,100){\line(1,0){20}}

\put(0,0){\vector(0,1){120}}
\put(20,0){\line(0,1){100}}
\put(40,0){\line(0,1){80}}
\put(60,0){\line(0,1){80}}
\put(80,0){\line(0,1){60}}
\put(100,0){\line(0,1){60}}
\put(120,0){\line(0,1){60}}
\put(140,0){\line(0,1){40}}
\put(160,0){\line(0,1){40}}
\put(180,0){\line(0,1){20}}
\put(200,0){\line(0,1){20}}

\put(-10,115){\mbox{$j$}}
\put(215,-10){\mbox{$i$}}

\put(48,28){\mbox{$\bullet$}}
\put(48,48){\mbox{$\bullet$}}
\put(48,68){\mbox{$\bullet$}}
\put(68,28){\mbox{$\bullet$}}
\put(88,28){\mbox{$\bullet$}}
\put(108,28){\mbox{$\bullet$}}
\put(128,28){\mbox{$\bullet$}}
\put(148,28){\mbox{$\bullet$}}

\end{picture}
\caption{{\footnotesize The Young diagram $\lambda=[5443332211]$ and the hook length $h_\Box=8$
for the box $\Box=(3,2)$ in it.}}
\label{picth}
\end{figure}
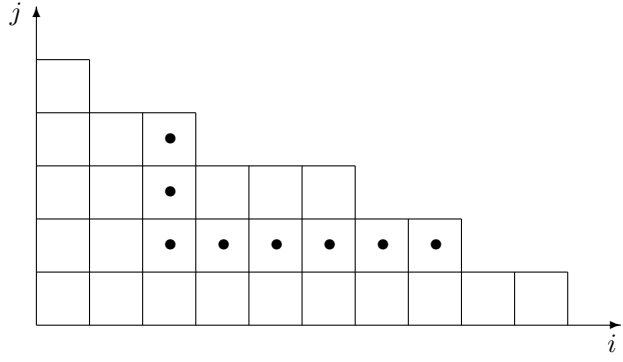

For rectangular representations $R=[r^s]$, the known figure-eight DE is
\cite{1605.09728}
\be
H^{4_1}_{[r^s]}=\sum_{\lambda\subseteq[r^s]}W_\lambda^{[r^s]}
\label{eq:rectangular-DE}
\ee
with weights
\be
W_\lambda^{[r^s]}=
\frac{\displaystyle\prod_{\Box\in\lambda}\Phi_{c_{_\Box}}^{r,s}}
{\displaystyle\prod_{\Box\in\lambda}[h_{_\Box}]^2}
\label{eq:rectangular-DE-weight}
\ee
with
\be
\Phi_c^{r,s}:=[r-c][s+c]D_{r+c}D_{c-s}.
\label{eq:rectangular-DE-input}
\ee
The empty diagram contributes one. For a box in row $i$ and column
$j$, $c_{_\Box}=j-i$ is its content, and $h_\Box$ is the number of
boxes in its hook, including the box itself; see Fig.~\ref{picth}.
The factors $\Phi_c^{r,s}$ contain the dependence on the color;
the hook lengths depend on the summation diagram $\lambda$.
At $s=1$, $\lambda=[k]$ has hook product $[k]!$, and this formula
reduces to (\ref{eq:symmetric-DE}).

First consider a summation diagram consisting of one hook,
$\lambda=[a,1^{b-1}]$. Its contents run from $1-b$ to $a-1$, and
its quantum hook product is
\[
[a+b-1][a-1]![b-1]!.
\]
Thus its weight is
\be
W_{[a,1^{b-1}]}^{[r^s]}
=
\underbrace{\frac{1}{[a+b-1][a-1]![b-1]!}}_{T_{ab}}\,
\underbrace{\frac{\displaystyle\prod_{c=1-b}^{a-1}\Phi_c^{r,s}}
{[a+b-1][a-1]![b-1]!}}_{Z_{ab}^{[r^s]}}.
\label{eq:hook-splitting}
\ee
One copy of the hook denominator is assigned to each matrix, and all
color factors are assigned to $Z$: the hook weight is $T_{ab}Z_{ab}$.
For a general $\lambda$, the corresponding minors reproduce the two
hook denominators and the content product, as shown in
Section~\ref{sec:41-minors}.

The normalization allows reciprocal row and column rescalings of
$T$ and $Z$ that preserve their paired minors. We choose $T$ to
depend only on $q$ and the hook indices, so that $Z$ retains the
color dependence when evolution is introduced.

\subsection{Rectangular representations}\label{sec:41-rect}

The splitting (\ref{eq:hook-splitting}) defines two $r\times s$
matrices with $a=1,\ldots,r$ and $b=1,\ldots,s$:
\be
Z^{[r^s]}_{ab} := \frac{1}{[a+b-1][a-1]![b-1]!}\prod_{c=1-b}^{a-1} [r-c][s+c]\{Aq^{r+c}\}\{Aq^{c-s}\}
\label{eq:Z-hook}
\ee
and
\be
T_{ab}:=\frac{1}{[a+b-1][a-1]![b-1]!}
\label{eq:T-hook}
\ee
Using the transpose $\tilde Z_{ba}:=Z_{ab}$, define square matrices
of sizes $r\times r$ and $s\times s$:
\be
R^{[r^s]}_{a,a'}:=   \delta_{a,a'} + \sum_{b=1}^s T_{ab}\tilde Z^{[r^s]}_{ba'}
\ee
and
\be
S^{[r^s]}_{b,b'}:=   \delta_{b,b'} + \sum_{a=1}^r \tilde Z^{[r^s]}_{ba}T_{ab'}
\ee
Then
\be
\boxed{
H^{4_1}_{[r^s]} = \det_{r\times r} R^{[r^s]} = \det_{s\times s} S^{[r^s]}
}
\label{41rectdet}
\ee
The two matrices have equal determinants; even when $r=s$, the
matrices themselves need not coincide or be transposes of one another.

For $s=1$, $T_{a1}=1/[a]!$ and
$Z_{a1}^{[r]}=[a]!\,U_a^{[r]}$. Hence the one-by-one determinant
is $1+\sum_aT_{a1}Z_{a1}=1+\sum_aU_a^{[r]}$, as required in (\ref{eq:symmetric-DE}).

\subsection{Why the determinant sums the required weights}
\label{sec:41-minors}

The proof has two steps. The quantum Cauchy formula evaluates the
minors of $T$ and identifies each DE weight with a product of minors
of $T$ and $Z$. Cauchy--Binet then sums these products into the
determinant (\ref{41rectdet}).

\paragraph{One diagram: the quantum Cauchy formula.}
Describe $\lambda$ by its diagonal hooks. For each of its $d$
diagonal boxes, let $a_i$ and $b_i$ be the arm and leg lengths,
respectively, with the diagonal box included in both.
Then
\[
a_1>\cdots>a_d\geq1,\qquad b_1>\cdots>b_d\geq1.
\]
For example, $\lambda=[2,1]$ has the single pair $(a_1,b_1)=(2,2)$,
whereas $\lambda=[2,2]$ has $(a_1,a_2)=(2,1)$ and
$(b_1,b_2)=(2,1)$. Thus the size of the minor is the number of
diagonal boxes, not the total number of boxes.

Write $\mathsf A=\{a_1,\ldots,a_d\}$ and
$\mathsf B=\{b_1,\ldots,b_d\}$. After removing the factorials
from rows and columns of $T_{\mathsf A,\mathsf B}$, the matrix
to evaluate has entries $1/[a_i+b_j-1]$. Its determinant is
\be
\det_{i,j}\frac{1}{[a_i+b_j-1]}
=
\frac{\displaystyle\prod_{i<j}[a_i-a_j][b_i-b_j]}
{\displaystyle\prod_{i,j}[a_i+b_j-1]}.
\label{eq:quantum-Cauchy}
\ee
This is the quantum Cauchy determinant formula. It follows from
the ordinary Cauchy determinant for $1/(x_i-y_j)$ by setting
$x_i=q^{2(a_i-1)}$, $y_j=q^{-2b_j}$, since
\[
\frac{1}{[a_i+b_j-1]}
=(q-q^{-1})\frac{q^{a_i-b_j-1}}{x_i-y_j}.
\]
Thus \emph{quantum} refers to the quantum-number entries; no new
summation rule is involved.

Restoring the factorials gives
\be
\det T_{\mathsf A,\mathsf B}
=
\frac{\displaystyle\prod_{i<j}[a_i-a_j][b_i-b_j]}
{\displaystyle\prod_i[a_i-1]![b_i-1]!
 \prod_{i,j}[a_i+b_j-1]}
=\frac{1}{\displaystyle\prod_{\Box\in\lambda}[h_{_\Box}]}.
\label{eq:T-minor}
\ee
The last equality is the hook-product formula in diagonal-hook
coordinates. Thus the minors of $T$ give the inverse hook product
for every diagram.

The color numerator separates into row and column factors. Put
\[
u_a=\prod_{c=0}^{a-1}\Phi_c^{r,s},\qquad
v_b=\prod_{c=1-b}^{-1}\Phi_c^{r,s}.
\]
Then $Z_{ab}=u_aT_{ab}v_b$, so
\be
\det Z_{\mathsf A,\mathsf B}
=
\left(\prod_i u_{a_i}v_{b_i}\right)
\det T_{\mathsf A,\mathsf B}
=
\left(\prod_{\Box\in\lambda}\Phi_{c_\Box}^{r,s}\right)
\det T_{\mathsf A,\mathsf B}.
\label{eq:Z-minor-factorization}
\ee
Here the diagonal hooks partition the boxes of $\lambda$, so their
content products multiply to the content product of the whole diagram.
Each weight in (\ref{eq:rectangular-DE}) is therefore a product of
two minors:
\be
\boxed{
W_\lambda^{[r^s]}
=\det T_{\mathsf A,\mathsf B}\det Z_{\mathsf A,\mathsf B}
=\frac{\displaystyle\prod_{\Box\in\lambda}\Phi_{c_{_\Box}}^{r,s}}
{\displaystyle\prod_{\Box\in\lambda}[h_{_\Box}]^2}.
}
\label{eq:W-minors}
\ee

\paragraph{All diagrams: Cauchy--Binet.}
For any two $r\times s$ matrices,
\be
\det(I_r+TZ^t)
=
\sum_{d=0}^{\min(r,s)}
\left( \sum_{\substack{\mathsf A\subseteq\{1,\ldots,r\},\
                  \mathsf B\subseteq\{1,\ldots,s\}\\
                  |\mathsf A|=|\mathsf B|=d}}
\det T_{\mathsf A,\mathsf B}\det Z_{\mathsf A,\mathsf B}
\right) =  \det(I_s+Z^tT)
\label{eq:CB-full}
\ee
To obtain this identity, first expand $\det(I_r+TZ^t)$ into principal
minors of $TZ^t$. Apply Cauchy--Binet to each minor of this matrix
product; its internal column set is $\mathsf B$. The empty minor is one.
In particular, the first terms are
\be
\det(I_r+TZ^t)
=1+\Tr T Z^t +\frac{1}{2}\Big((\Tr T Z^t)^2-  \Tr T Z^t  T Z^t\Big) + \ldots
=\nn \\
=  1+\sum_{a,b}T_{ab}Z_{ab} +\sum_{a<a',\,b<b'}
(T_{ab}T_{a'b'}-T_{ab'}T_{a'b})
(Z_{ab}Z_{a'b'}-Z_{ab'}Z_{a'b})+\cdots .
\label{eq:CB-expanded}
\ee
The first line is the usual expansion $\det(1+A) = \exp(\Tr\log(1+A))$,
while the second line expresses it as a sum of paired minors.
Each pair of equally sized index sets specifies exactly one diagram
$\lambda\subseteq[r^s]$ through its diagonal hooks. By
(\ref{eq:W-minors}), every summand is therefore the required
$W_\lambda^{[r^s]}$. This proves (\ref{41rectdet}).
Expanding $\det(I_s+Z^tT)$ gives the same sum.

The cross terms in (\ref{eq:CB-expanded}) show why the formula uses
a matrix product: taking a determinant of the entry-wise products
$T_{ab}Z_{ab}$ would omit them.

\paragraph{The complete example $R=[2,2]=(2,1|2,1)$.}
There are six sub-diagrams $\lambda=(\vec a|\vec b)$. Their weights are:
\[
\begin{array}{c|c|c|c}
\lambda&d& (\vec a|\vec b)& W_\lambda^{[2,2]}\\ \hline
\emptyset&0&(\ |\ )&1\\
{[1]}&1&(1|1)& T_{11}Z_{11}=\Phi_0\\
{[2]}&1&(2|1)& T_{21}Z_{21}=\Phi_0\Phi_1/[2]^2\\
{[1,1]}&1&(1|2)&T_{12}Z_{12}=\Phi_{-1}\Phi_0/[2]^2\\
{[2,1]}&1&(2|2)&T_{22}Z_{22}=\Phi_{-1}\Phi_0\Phi_1/[3]^2\\
{[2,2]}&2&(2,1|2,1) &\det T\det Z
=\Phi_{-1}\Phi_0^2\Phi_1/([3]^2[2]^4)
\end{array}
\]
Here $\Phi_c=\Phi_c^{2,2}$. The last denominator can be checked directly:
\[
T=\begin{pmatrix} T_{11} & T_{12} \\ T_{21} & T_{22}   \end{pmatrix}
=\begin{pmatrix}1&\frac{1}{[2]}\\\frac{1}{[2]}&\frac{1}{[3]}\end{pmatrix},
\qquad
\det T=\frac{1}{[3]}-\frac{1}{[2]^2}
=\frac{1}{[3][2]^2},
\]
using $[2]^2-[3]=1$. The hook lengths of $[2,2]$ are $3,2,2,1$,
so this is exactly their inverse quantum product.
Likewise,
\be
Z=\begin{pmatrix} Z_{11} & Z_{12} \\ Z_{21} & Z_{22}   \end{pmatrix}
=\begin{pmatrix}\Phi_0&\frac{\Phi_{-1}\Phi_0}{[2]}\\\frac{\Phi_{0}\Phi_1}{[2]}&\frac{\Phi_{-1}\Phi_0\Phi_1}{[3]}\end{pmatrix},
\qquad
\det Z
=\frac{\Phi_{-1}\Phi_0^2\Phi_1}{[3][2]^2},
\label{eq:Z41-22-expansion}
\ee
Adding the six rows gives
\be
H^{4_1}_{[2,2]}
=1+T_{11}Z_{11}+T_{12}Z_{12}+T_{21}Z_{21}+T_{22}Z_{22}
+\det T\det Z.
\label{eq:41-22-expansion}
\ee
In particular, the one-entry term $T_{22}Z_{22}$ describes the
three-box hook $[2,1]$, and the two-by-two minors describe the
four-box square $[2,2]$.

\subsection{A balanced form and the root-of-unity limit}
\label{sec:41-cyclotomic}

The matrices $T$ and $Z$ were introduced asymmetrically: $T$ is universal,
whereas all color dependence is placed in $Z$.  For the determinant this
asymmetry is convenient, but for a root-of-unity limit it hides a simple
fact: the differential-expansion weights only involve \emph{paired} minors
of $T$ and $Z$.  We first put these two minors on equal footing.

Recall from (\ref{eq:Z-minor-factorization}) that
\[
 Z=UTV,\qquad
 U=\operatorname{diag}(u_1,\ldots,u_r),\qquad
 V=\operatorname{diag}(v_1,\ldots,v_s).
\]
At generic parameters choose local square roots of the diagonal factors and
set
\be
 B:=U^{1/2}TV^{1/2}.
 \label{eq:balanced-B}
\ee
This introduces no new information: it is only a symmetric gauge for the
pair $(T,Z)$.  Indeed,
\be
 \boxed{
 H^{4_1}_{[r^s]}
 =\det_r(I_r+BB^t)
 =\det_s(I_s+B^tB),
 }
 \label{eq:balanced-det}
\ee
and for the minor corresponding to a diagram $\lambda$,
\be
 \boxed{
 \left(\det B_{\mathsf A,\mathsf B}\right)^2
 =\det T_{\mathsf A,\mathsf B}\det Z_{\mathsf A,\mathsf B}
 =W_\lambda^{[r^s]}.
 }
 \label{eq:balanced-minors}
\ee
Thus $B$ is useful because every DE weight is the square of a single
Plucker coordinate.  The choice of square-root branches drops out of all
formulas for the polynomial.

\paragraph{A motivating example.}
For $R=[2,2]$ write $x_c=D_{2+c}D_{c-2}$.  The six terms displayed above
combine into
\[
 H^{4_1}_{[2,2]}
 =1+[2]^2x_0+[3]x_0(x_{-1}+x_1)
   +[2]^2x_{-1}x_0x_1+x_{-1}x_0^2x_1.
\]
At $Q=q^2$ with $Q^3=1$, one has $[3]=0$ and $[2]^2=1$, hence
\be
 \boxed{
 H^{4_1}_{[2,2]}\big|_{Q^3=1}
 =(1+x_0)(1+x_{-1}x_0x_1).
 }
 \label{eq:41-22-root-factorization}
\ee
We now show that this elementary factorization is the first example of a
general block decomposition of $B$.

\paragraph{Where the zeros occur.}
Let $Q_0$ be a primitive $\ell$-th root of unity,
\[
 Q=q^2,\qquad Q_0^\ell=1,
\]
and in this paragraph assume
\[
 1\le r,s<\ell,
\]
with $A$ kept generic.  Then $[n]=0$ exactly when $n$ is divisible by
$\ell$.  If $r+s\le\ell$, all quantum numbers in the hook denominators
satisfy $1\le a+b-1<\ell$, so nothing singular happens.  The first new
case is therefore
\[
 d:=r+s-\ell>0.
\]
It is convenient to put
\[
 \bar r:=\ell-s,\qquad \bar s:=\ell-r,
 \qquad r=\bar r+d,\quad s=\bar s+d.
\]

For a one-hook diagram $[a,1^{b-1}]$, the one-by-one version of
(\ref{eq:balanced-minors}) is
\be
 B_{ab}^{\,2}
 =\frac{\displaystyle\prod_{c=1-b}^{a-1}
 [r-c][s+c]D_{r+c}D_{c-s}}
 {[a+b-1]^2[a-1]!^2[b-1]!^2}.
 \label{eq:B-hook-square}
\ee
Since $a,b<\ell$, the factorials are nonzero at $Q=Q_0$.  The only issue
is therefore the content interval
\[
 1-b\le c\le a-1.
\]
There are two possible numerator zeros inside this interval:
\[
 [s+c]=0\quad\hbox{at}\quad c=\ell-s=\bar r,
 \qquad
 [r-c]=0\quad\hbox{at}\quad c=r-\ell=-\bar s.
\]
Consequently:
\begin{itemize}
\item if $a\le\bar r$ and $b\le\bar s$, neither zero is present and
      $B_{ab}$ has a regular generic limit;
\item if the hook leaves the $\bar r\times\bar s$ corner but
      $a+b-1\ne\ell$, the numerator has an uncancelled zero and
      $B_{ab}$ vanishes;
\item if $a+b-1=\ell$, both numerator zeros are present and cancel the
      squared zero $[a+b-1]^2=[\ell]^2$ in the denominator.
\end{itemize}
Thus the surviving entries have the simple pattern
\be
 B_{ab}(Q_0)\ne0\quad\Longrightarrow\quad
 \left\{
 \begin{array}{ll}
 a\le\bar r,\ b\le\bar s,&\text{the residual rectangle},\\[1mm]
 a+b-1=\ell,&\text{one length-$\ell$ hook}.
 \end{array}
 \right.
 \label{eq:B-cyclotomic-support}
\ee

This pattern is easier to see as a matrix.  Split the rows as
$\bar r+d$ and the columns as $\bar s+d$.  The off-diagonal rectangular
blocks vanish, while the lower-right $d\times d$ block has nonzero entries
only on its anti-diagonal.  Reversing these last $d$ columns therefore gives
\be
 \boxed{
 B(Q_0)P
 =B_{\rm core}\oplus
 \operatorname{diag}(\beta_1,\ldots,\beta_d),
 }
 \label{eq:B-cyclotomic-block}
\ee
where $P$ is this column reversal, $B_{\rm core}$ is the
$\bar r\times\bar s$ north-west block, and
\be
 \beta_\alpha^2
 :=\lim_{Q\to Q_0}
 W_{[\bar r+\alpha,\,1^{\,s-\alpha}]}^{[r^s]},
 \qquad \alpha=1,\ldots,d.
 \label{eq:beta-cyclotomic}
\ee
The hook in the last formula has
$a+b-1=\ell$, hence total hook length $\ell$.

Now the factorization of the knot polynomial is simply the determinant of a
block-diagonal matrix:
\be
 \boxed{
 H^{4_1}_{[r^s]}(A,q_0)
 =\det(I_{\bar r}+B_{\rm core}B_{\rm core}^{t})
 \prod_{\alpha=1}^{d}(1+\beta_\alpha^2).
 }
 \label{eq:41-cyclotomic-factorization}
\ee
The first factor is a residual block of the original color; it should not
in general be identified with the HOMFLY polynomial in the smaller color
$[\bar r^{\bar s}]$, because its diagonal dressing is still inherited from
$(r,s)$.  What is universal is the splitting into one residual rectangular
sector and $d=r+s-\ell$ independent length-$\ell$ hook sectors.

\subsection{Non-rectangular representations: what changes for $[2,1]$}
\label{sec:41-nonrect}

We now take the color $R=[2,1]$. Its polynomial requires its own
color weights: deleting the last term of (\ref{eq:41-22-expansion})
would leave the factors $\Phi_c^{2,2}$ of the square.
Both polynomials can nevertheless be compared through expansions
in differential pairs. In the following $[2,1]$ formula, the terms
are indexed by all subsets of its three boxes.

Write $z_{_{I|J}}=D_I D_{-J}$ and $\epsilon=q-q^{-1}$.
Lower-case $z$ denotes one differential pair, whereas $Z_{ab}$
above denotes an entire hook entry, with its quantum-number factors.
The known exact polynomial is \cite{1211.6375}
\begin{align}
H^{4_1}_{[2,1]}={}&1
+\underbrace{z_{3|3}+z_{2|0}+z_{0|2}}_{\text{one selected box}}
\nonumber\\
&+\underbrace{z_{4|2}z_{2|0}+z_{2|4}z_{0|2}
+(1-\epsilon^2)z_{2|0}z_{0|2}}_{\text{two selected boxes}}
\nonumber\\
&+\underbrace{z_{3|3}z_{2|0}z_{0|2}}_{\text{three selected boxes}}
\label{eq:41-21}
\end{align}
To specify the correspondence, label the corner, right and upper
boxes
by $C,H,V$:
\[
\begin{array}{c}
\boxed{V}\phantom{\boxed{H}}\\
\boxed{C}\boxed{H}
\end{array}
\qquad
C=(1,1),\quad H=(1,2),\quad V=(2,1).
\]
In the shifted-box convention of \cite{1211.6375}, the eight terms are
\[
\begin{array}{c|l|l}
\text{selected subset}&\text{geometry}&\text{contribution}\\ \hline
\emptyset&\text{no boxes}&1\\
\{C\}&\text{corner}&z_{3|3}\\
\{H\}&\text{right box}&z_{0|2}\\
\{V\}&\text{upper box}&z_{2|0}\\
\{C,H\}&\text{horizontal pair}&z_{2|4}z_{0|2}\\
\{C,V\}&\text{vertical pair}&z_{4|2}z_{2|0}\\
\{H,V\}&\text{disconnected pair}&(1-\epsilon^2)z_{2|0}z_{0|2}\\
\{C,H,V\}&\text{whole hook}&z_{3|3}z_{2|0}z_{0|2}
\end{array}
\]
The corner factor becomes $z_{2|4}$ for $\{C,H\}$ and $z_{4|2}$
for $\{C,V\}$, while the disconnected pair carries the coefficient
$1-\epsilon^2$. These corrections to independent box weights are
explicit in the equivalent formula
\begin{align*}
H^{4_1}_{[2,1]}={}&
(1+z_{3|3})(1+z_{0|2})(1+z_{2|0})\\
&+(z_{2|4}-z_{3|3})z_{0|2}
 +(z_{4|2}-z_{3|3})z_{2|0}
 -\epsilon^2 z_{2|0}z_{0|2}.
\end{align*}
The eight box subsets distinguish positions in the color diagram.
They differ from its five Young sub-diagrams,
$\emptyset,[1],[2],[1,1],[2,1]$, since the isolated boxes $H,V$ and
the pair $\{H,V\}$ are not anchored at the corner. The regrouping
into the six evolution channels is described in
Section~\ref{sec:twist-nonrect}.

For the rectangular color $[2,2]$, a corresponding box-subset expansion is
\be
H^{4_1}_{[2,2]} = &1+[2]^2z_{2|2}
 +[3]z_{2|2}(z_{3|1}+z_{1|3})
 +[2]^2z_{2|2}z_{3|1}z_{1|3}
 +z_{2|2}^{\,2}z_{3|1}z_{1|3} = \nn \\ \nn \\
&=1
+\underbrace{z_{3|3}+z_{3|1}+z_{1|3}+z_{1|1}}_{\text{one selected box - 4 combinations}}
\nonumber\\
&+\underbrace{   z_{3|1}(z_{4|2}+z_{2|2}+z_{0|2}) +z_{1|3}(z_{2|4}+z_{2|2}+z_{2|0})
}_{\text{two selected boxes - 6 combinations}}
\nonumber\\
&+\underbrace{z_{3|1}z_{1|3}(z_{3|3}+z_{3|1}+z_{1|3}+z_{1|1})}_{\text{three selected boxes - 4 combinations}}
\nonumber\\
&+\underbrace{z_{3|3}z_{2|2}^2z_{1|1}}_{\text{four selected boxes - 1 combination}}
\label{eq:41-22}
\ee
The DE (\ref{eq:rectangular-DE}) groups terms by Young sub-diagrams;
it does not specify a unique box-subset decomposition. For example,
the quadratic contribution in (\ref{eq:41-22}) can also be written as
$(z_{5|1}+z_{3|1}+2z_{1|1}+z_{1|3}+z_{1|5})z_{2|2}$.
Such decompositions allow comparison with non-rectangular colors,
but do not by themselves supply the paired-minor structure used above.
The $\epsilon$-dependent correction in (\ref{eq:41-21}) must be
included in that structure; its presence alone does not rule out
a determinant representation.

The distinction between positions and grouped weights is already
visible at $q=1$. Put $z=\{A\}^2$. Every $z_{I|J}$ becomes $z$,
the shifts disappear, and (\ref{eq:41-21}) gives
\[
H^{4_1}_{[2,1]}(A,1)=1+3z+3z^2+z^3=(1+z)^3.
\]
For the rectangle $[2,2]$, the six weights listed in
Section~\ref{sec:41-minors} become
$1,4z,3z^2,3z^2,4z^3,z^4$, whose sum is $(1+z)^4$.
Thus the rectangular weight of $[1]$ already combines four box
positions. A non-rectangular counterpart of (\ref{41rectdet}) would
require regrouping the exact weights, including the shifted pairs
and the disconnected coefficient, into products of compatible minors.
For rectangles this is ensured by $Z_{ab}=u_aT_{ab}v_b$ with the
universal content factor $\Phi_c^{r,s}$. The corresponding
non-rectangular factors remain to be found.

\section{Twist knots}\label{sec:twists}

We use the twist parameter $m$ with
${\rm tw}_{-1}=4_1$, ${\rm tw}_0=\text{unknot}$ and
${\rm tw}_1=3_1$. The color matrix $Z$ stays exactly as in
(\ref{eq:Z-hook}), while $T$ is replaced by the hook evolution
matrix $T^{(m)}$. Its construction follows the evolution approach
\cite{evo} and the KNTZ formulas \cite{KNTZ,1902.04140}.
We first give its entries and the resulting determinant, then derive
the resummation of the differential expansion.

\subsection{Symmetric representations}

For $s=1$ the matrix $T^{(m)}$ has one column. Write it as
$t_a^{(m)}=T_{a1}^{(m)}$, $1\leq a\leq r$. Its entries are \cite{evo}
\be
\boxed{
t_a^{(m)}=A^a q^{a(a-1)/2}
\sum_{i=1}^{a}
\frac{(-1)^{i-1}D_{2i-1}
\left(1-A^{2mi}q^{2mi(i-1)}\right)}
{[i]![a-i]!\displaystyle\prod_{k=i-1}^{a+i-1}D_k}.
}
\label{eq:t-symmetric}
\ee
These are finite sums of explicit rational functions, with all
$m$-dependence in the displayed powers. This is the usual symmetric
twist evolution written in the normalization needed by the determinant
\cite{1606.06015}. In particular,
\begin{align}
t_1^{(m)}&=\frac{A(1-A^{2m})}{D_0},\nonumber\\
t_2^{(m)}&=A^2q\left(
\frac{1-A^{2m}}{D_0D_2}
-\frac{1-A^{4m}q^{4m}}{[2]D_1D_2}\right).
\label{eq:t-first-two}
\end{align}
The first column of the color matrix is also explicit:
\be
Z_{a1}^{[r]}=\frac{[r]!}{[r-a]!}
\prod_{k=0}^{a-1}D_{r+k}D_{k-1}.
\ee
Thus the analogue of (\ref{41rdet}) is
\be
\boxed{
H_{[r]}^{{\rm tw}_m}
=\det_{1\leq a,a'\leq r}
\left(\delta_{aa'}+t_a^{(m)}Z_{a'1}^{[r]}\right)
=1+\sum_{a=1}^r t_a^{(m)}Z_{a1}^{[r]}.
}
\label{eq:twist-symmetric}
\ee
At $m=-1$, $t_a^{(-1)}=1/[a]!$, and the last sum is precisely
the figure-eight sum in (\ref{41rdet}). For $[1]$ this gives the
particularly simple answer
\be
H_{[1]}^{{\rm tw}_m}=1+D_1D_{-1}\frac{A(1-A^{2m})}{D_0}.
\ee

\subsection{From a column to the explicit hook matrix}

For a rectangle we need $1\leq a\leq r$, $1\leq b\leq s$.
Put
\be
\lambda_{ij}:=(Aq^{i-j})^{2(i+j-1)},\qquad
c_{ab}:=\frac{a(a-1)-b(b-1)}2.
\label{eq:hook-eigenvalues}
\ee
The symbol $\lambda_{ij}$ here denotes an evolution eigenvalue;
its indices label the hook $[i,1^{j-1}]$.
For $1\leq i\leq a$ and $1\leq j\leq b$ define
\be
C_{ab}^{ij}:=
\frac{(-1)^{i+j}D_{2i-1}D_{1-2j}}
{[i-1]![a-i]![j-1]![b-j]![i+j-1]D_{i-j}
\displaystyle\prod_{k=i}^{a+i-1}D_k
\displaystyle\prod_{k=j}^{b+j-1}D_{-k}}.
\label{eq:explicit-hook-coefficients}
\ee
Then the required $r\times s$ matrix is
\be
\boxed{
T_{ab}^{(m)}=A^{a+b-1}q^{c_{ab}}
\sum_{i=1}^{a}\sum_{j=1}^{b} C_{ab}^{ij}
\left(1-\lambda_{ij}^{m}\right).
}
\label{eq:explicit-Tm}
\ee
This form of the single-hook evolution follows from the finite-sum
formula in \cite[Eq.~(89)]{1606.06015}; the constant term has been
combined with the spectral terms by using its value at $m=0$.
The entries do not depend on $r,s$; the rectangle only bounds their
indices. Setting $b=1$ in (\ref{eq:explicit-hook-coefficients}) cancels
$D_{-1}$ and gives exactly (\ref{eq:t-symmetric}).

Three normalizations are useful to keep in mind:
\be
T_{ab}^{(-1)}=T_{ab},\qquad T_{ab}^{(0)}=0,\qquad
T_{ab}^{(1)}=
\frac{(-A^2)^{a+b-1}q^{a(a-1)-b(b-1)}}
{[a+b-1][a-1]![b-1]!}.
\label{eq:T-three-knots}
\ee
The apparent differential denominators in individual summands of
(\ref{eq:explicit-Tm}) cancel in the finite evolution coefficient.
The expressions are understood as defined at generic $A,q$; at a removable pole
one first combines the terms.

\subsection{Rectangular representations}\label{sec:twist-rect}

As in Section~\ref{sec:41-rect}, define two square matrices:
\be
R^{[r^s];(m)}_{aa'}:=\delta_{aa'}+
\sum_{b=1}^s T_{ab}^{(m)}\tilde Z^{[r^s]}_{ba'},\qquad
S^{[r^s];(m)}_{bb'}:=\delta_{bb'}+
\sum_{a=1}^r\tilde Z^{[r^s]}_{ba}T_{ab'}^{(m)}.
\ee
Every entry on the right is given by the products in
(\ref{eq:Z-hook}) and the finite sum (\ref{eq:explicit-Tm}). The answer is
\be
\boxed{
H_{[r^s]}^{{\rm tw}_m}
=\det_{r\times r}R^{[r^s];(m)}
=\det_{s\times s}S^{[r^s];(m)}.
}
\label{eq:twist-rectdet}
\ee
The finite sums enter the matrix entries, while the sum over
intermediate diagrams is collected into a determinant of order
$\min(r,s)$.

\subsection{The first square example and the resummation}

For $r=s=2$, the hook evolution matrix has four entries:
\begin{align}
T_{11}^{(m)}={}&\frac{A(1-A^{2m})}{D_0},\nonumber\\
T_{21}^{(m)}={}&A^2q\left(
\frac{1-A^{2m}}{D_0D_2}
-\frac{1-A^{4m}q^{4m}}{[2]D_1D_2}\right),\nonumber\\
T_{12}^{(m)}={}&A^2q^{-1}\left(
\frac{1-A^{2m}}{D_0D_{-2}}
-\frac{1-A^{4m}q^{-4m}}{[2]D_{-1}D_{-2}}\right),\nonumber\\
T_{22}^{(m)}={}&\frac{A^3}{D_2D_{-2}}\left(
\frac{1-A^{2m}}{D_0}
-\frac{1-A^{4m}q^{4m}}{[2]D_1}
-\frac{1-A^{4m}q^{-4m}}{[2]D_{-1}}
+\frac{1-A^{6m}}{[3]D_0}\right).
\label{eq:T22-explicit}
\end{align}
The color matrix is still exactly the same as in (\ref{eq:Z41-22-expansion}):
\be
Z^{[2,2]}=[2]^2D_2D_{-2}
\begin{pmatrix}
1&\dfrac{[3]}{[2]}D_1D_{-3}\\[2mm]
\dfrac{[3]}{[2]}D_3D_{-1}&[3]D_1D_{-3}D_3D_{-1}
\end{pmatrix}.
\label{eq:Z22-explicit}
\ee
What changes with $m$ is (\ref{eq:41-22-expansion}).
Substitution into (\ref{eq:twist-rectdet}) gives an explicit
$2\times2$ determinant for every integer $m$.
For example,
\be
T^{(-1)}=\begin{pmatrix}1&1/[2]\\1/[2]&1/[3]\end{pmatrix},\qquad
T^{(1)}=\begin{pmatrix}-A^2&A^4q^{-2}/[2]\\
A^4q^2/[2]&-A^6/[3]\end{pmatrix}.
\ee
where $T^{(-1)}$ is just (\ref{eq:41-22-expansion}), but already $T^{(1)}$ is different.

Now the DE states that
\be
H^{{\rm tw}_m}_{[r^s]}=\sum_\lambda W^{[r^s]}_\lambda F_\lambda^{(m)}
\ee
The main point is that {\bf $F^{(m)}_\lambda$ is still a minor}!

Indeed, the exterior-power form of KNTZ evolution, derived in
Appendix~\ref{app:matrix-origin}, gives the following identity for
the hook sets $\mathsf A,\mathsf B$ of Section~\ref{sec:41-minors}:
\be
\boxed{
F_\lambda^{(m)}=
\frac{\det T^{(m)}_{\mathsf A,\mathsf B}}
{\det T_{\mathsf A,\mathsf B}}.
\label{eq:F-minor-ratio}
}
\ee
The denominator is the inverse quantum hook product in
(\ref{eq:T-minor}), which is nonzero at generic $q$.
Multiplication by (\ref{eq:W-minors}) cancels it:
\be
W^{[r^s]}_\lambda F_\lambda^{(m)}
=\det Z^{[r^s]}_{\mathsf A,\mathsf B}
\det T^{(m)}_{\mathsf A,\mathsf B}.
\label{eq:twist-minor-term}
\ee
Summing these terms by (\ref{eq:CB-full}) proves
(\ref{eq:twist-rectdet}). For $[2,2]$, the four hook terms are
$Z_{ab}T_{ab}^{(m)}$ and the square contributes
$\det Z\det T^{(m)}$.

\subsection{Non-rectangular representations}\label{sec:twist-nonrect}

For the non-rectangular color $[2,1]$, the reduced evolution
requires an additional channel beyond the five Young sub-diagrams.
The six channels and their eigenvalues are
\be
\emptyset,\ [1],\ [1,1],\ [2],\ [2,1],\ X_2,
\qquad
\Lambda=\operatorname{diag}(1,A^2,A^4q^{-4},A^4q^4,A^6,A^4),
\label{eq:R21-spectrum}
\ee
where $X_2=([2],[1,1])\oplus([1,1],[2])$ combines two conjugate
off-diagonal channels with the same eigenvalue \cite{1903.00259}.
Their eigenvalue $A^4$ does not occur in the single-hook
coefficients above.

This six-channel system cannot be the exterior square of a
four-level evolution, even up to a common scalar normalization.
Indeed, the six pairwise products of four one-particle eigenvalues
can be grouped into three complementary pairs with equal product.
For (\ref{eq:R21-spectrum}), the total $A$-degree is $20$; the
common degree of such three pair-products would have to be $20/3$,
whereas each pair has integral degree. Thus the same four-level
mechanism as in $[2,2]$ cannot describe the whole reduced $[2,1]$
evolution. A determinant with additional matrix data remains possible,
but is not provided by the rectangular formulas.

The DE coefficients for these channels are given in
\cite[Eqs.~(11)--(14)]{1903.00259}. Denote them by $\cZ$ to
distinguish them from our matrix entries $Z_{ab}$, and put
$P=D_2D_{-2}$. In the channel notation above, they are
\[
\begin{aligned}
 \cZ_{[1]}&=\frac{[3]D_0^2+[3]^2P}{[2]^2},\\
 \cZ_{[2]}&=\frac{[3]}{[2]}PD_0D_3,
 &\cZ_{[1,1]}&=\frac{[3]}{[2]}PD_0D_{-3},\\
 \cZ_{[2,1]}&=PD_3D_{-3}D_1D_{-1},
 &\cZ_{X_2}&=-[3]^2\epsilon^4P.
\end{aligned}
\]
For $4_1$, the corresponding evolution factors are all one, so
\[
 H^{4_1}_{[2,1]}=1+\cZ_{[1]}+\cZ_{[2]}+\cZ_{[1,1]}
                    +\cZ_{[2,1]}+\cZ_{X_2}.
\]

To relate these coefficients to (\ref{eq:41-21}), denote its three nonconstant lines by
\[
\begin{aligned}
 U_1&=z_{3|3}+z_{2|0}+z_{0|2},\\
 U_2&=z_{4|2}z_{2|0}+z_{2|4}z_{0|2}
                     +(1-\epsilon^2)z_{2|0}z_{0|2},\\
 U_3&=z_{3|3}z_{2|0}z_{0|2}.
\end{aligned}
\]
Direct regrouping gives
\[
\boxed{\begin{aligned}
 U_1&=\cZ_{[1]},\\
 U_2&=\cZ_{[2]}+\cZ_{[1,1]}-\epsilon^2PD_0^2,\\
 U_3&=\cZ_{[2,1]}+\epsilon^2PD_3D_{-3}.
\end{aligned}}
\]
For the second equality use
\[
 D_k+D_{-k}=(q^k+q^{-k})D_0,\qquad
 \frac{[3]}{[2]}(q^3+q^{-3})=q^4+1+q^{-4}.
\]
For the third, use $D_1D_{-1}=D_0^2-\epsilon^2$.
Finally, the two remainders combine into exactly the extra coefficient:
\[
\begin{aligned}
 -\epsilon^2PD_0^2+\epsilon^2PD_3D_{-3}
 &=\epsilon^2P(D_3D_{-3}-D_0^2)\\
 &=-[3]^2\epsilon^4P=\cZ_{X_2},
\end{aligned}
\]
since $D_kD_{-k}=D_0^2-\epsilon^2[k]^2$.
Thus $X_2$ collects the remainders from both the quadratic
and cubic lines of (\ref{eq:41-21}). It is not just the disconnected-pair correction
$-\epsilon^2z_{2|0}z_{0|2}$ taken separately.

\paragraph{A Schur decomposition beyond the first anomalous channel.}
The first genuinely new off-diagonal channel for the color $[3,2]$ is

$$
X_{31,22}=([3,1],[2,2])\oplus([2,2],[3,1]).
$$

Its evolution eigenvalue is $A^8q^4$, while the diagonal channel $[3,2]$
has eigenvalue $A^{10}q^8$.  The difference of the corresponding
evolution functions has the regular factor $A^2q^4-1$.  We therefore put

$$
G_{31,22}^{(m)}
:=
\frac{F_{X_{31,22}}^{(m)}-F_{[3,2]}^{(m)}}{A^2q^4-1}.
$$

Remarkably, this regularized function is still expressible in terms of
ordinary Schur polynomials, although a single character is no longer
sufficient.  Set

$$
x=A^2,\qquad z=q^2,\qquad n=m-2
$$

and define

$$
\begin{aligned}
S_0(n)&=
s_{[n,n]}(Aq^{-2},Aq^2,A^3q^{-2},A^3q^2),\\
S_1(n)&=
s_{[n,n]}(A^2q^{-1},A^2q^5,A^4q,A^4q^7),\\
S_2(n)&=
s_{[n,n]}(A^3q^{-2},A^3q^2,A^5q^2,A^5q^6).
\end{aligned}
$$

Then, for all $m\ge2$,
\be
\boxed{
\begin{aligned}
G_{31,22}^{(m)}
={}&
\frac{x^7z^2}{(xz-1)(xz^3-1)} \cdot S_0(m-2)-\frac{x^9z^7(z^2+1)}{(x-z)(xz^3-1)} \cdot S_1(m-2)+\frac{x^{11}z^5}{(x-z)(xz-1)} \cdot S_2(m-2).
\end{aligned}}
\label{eq:G3122-Schur}
\ee
The coefficients in this decomposition are independent of $m$.
Although the three terms on the right-hand side have apparent poles,
their sum is regular.

The identity can be checked spectrally.  Each character
$s_{[n,n]}$ above contains five evolution frequencies; their union is
precisely the eleven non-vacuum frequencies of the reduced $[3,2]$
system.  Equality of the corresponding eleven spectral coefficients
therefore proves (\ref{eq:G3122-Schur}) for all $m\ge2$.
As an additional check, direct evaluation of the original evolution
matrix agrees for $m=2,\ldots,6$.

At $q=1$ the three sets of Schur variables become proportional, and
(\ref{eq:G3122-Schur}) reduces to

$$
G_{31,22}^{(m)}(A,1)
=
A^4
\left(\frac{A^{2m}-1}{A^2-1}\right)^2
G_2^{(m)}(A,1).
$$

Thus the previously observed factor $m^2$ at $A\to1$ is the degenerate
limit of this three-character decomposition.  This example suggests
that, although the single exterior-power description does not extend
directly to non-rectangular colors, parts of the anomalous evolution may
still admit finite decompositions into ordinary Schur characters.

\section{Double braids}\label{sec:double-braids}

For antiparallel double braids the two twist evolutions factorize
channel by channel \cite{1606.06015}:
\be
F_\lambda^{(m,n)}=
\frac{F_\lambda^{(m)}F_\lambda^{(n)}}{F_\lambda^{(1)}}.
\label{eq:double-factorization}
\ee
We choose parameters so that ${\rm db}_{m,1}={\rm tw}_m$ in this
normalization. The division by the trefoil coefficient is essential.
Its special product form is what makes the final determinant possible.

\subsection{Symmetric representations}

Keep the explicit column $t_a^{(m)}$ of (\ref{eq:t-symmetric}), and put
\be
\ell_a^{[r]}=
\frac{[r]![a]!}{[r-a]!\,(-A^2)^a q^{a(a-1)}}
\prod_{k=0}^{a-1}D_{r+k}D_{k-1}.
\ee
Then
\be
\boxed{
H_{[r]}^{{\rm db}_{m,n}}
=\det_{1\leq a,a'\leq r}
\left(\delta_{aa'}+t_a^{(m)}t_{a'}^{(n)}\ell_{a'}^{[r]}\right)
=1+\sum_{a=1}^r\ell_a^{[r]}t_a^{(m)}t_a^{(n)}.
}
\label{eq:double-symmetric}
\ee
For the fundamental color this reduces to
\be
\boxed{
H_{[1]}^{{\rm db}_{m,n}}
=1-D_1D_{-1}\frac{(1-A^{2m})(1-A^{2n})}{D_0^2}.
}
\label{eq:double-fundamental}
\ee
At $n=1$ this becomes the fundamental twist formula above.
At $m=0$ or $n=0$ it equals one.

\subsection{The two diagonal factors}

As we already saw in (\ref{eq:Z-minor-factorization}),
the color matrix \(Z\) differs from the reference matrix \(T\) only by diagonal row and column factors.
For double braids, the crucial observation is that an analogous factorization holds
with the trefoil evolution matrix \(T^{(1)}\).
This is precisely what allows the trefoil reference minors in the denominator of (\ref{eq:double-factorization})
to cancel and makes the Cauchy--Binet resummation possible.

For a rectangle, recall the color factors
\be
\Phi_c^{r,s}=[r-c][s+c]D_{r+c}D_{c-s}.
\ee
Define an $r\times r$ diagonal matrix $L$ and an $s\times s$
diagonal matrix $M$ by
\begin{align}
L_{aa}&=\ell_a:=
\frac{\displaystyle\prod_{c=0}^{a-1}\Phi_c^{r,s}}
{(-A^2)^a q^{a(a-1)}},\nonumber\\
M_{bb}&=\mu_b:=
\frac{q^{b(b-1)}}{(-A^2)^{b-1}}
\prod_{c=1-b}^{-1}\Phi_c^{r,s}.
\label{eq:LM-products}
\end{align}
In particular $\mu_1=1$. Using the explicit trefoil entry in
(\ref{eq:T-three-knots}), one checks directly that
\be
Z_{ab}^{[r^s]}=\ell_a T_{ab}^{(1)}\mu_b,
\qquad Z=L T^{(1)}M.
\label{eq:Z-trefoil-factorization}
\ee
The ratio of the color entry to the trefoil entry thus separates
into a row factor and a column factor. This separation allows the
reference minors to cancel in the double-braid sum.

\subsection{Rectangular representations}\label{sec:double-rect}

The explicit matrices replacing those of Section~\ref{sec:41-rect} are
\begin{align}
R^{[r^s];(m,n)}_{aa'}&:=\delta_{aa'}+
\ell_{a'}\sum_{b=1}^s\mu_b T_{ab}^{(m)}T_{a'b}^{(n)},\nonumber\\
S^{[r^s];(m,n)}_{bb'}&:=\delta_{bb'}+
\mu_b\sum_{a=1}^r\ell_a T_{ab}^{(n)}T_{ab'}^{(m)}.
\label{eq:double-matrix-entries}
\end{align}
Here $T^{(m)}$ and $T^{(n)}$ are the finite sums
(\ref{eq:explicit-Tm}), and $\ell_a,\mu_b$ are the products
(\ref{eq:LM-products}). The resulting determinant is
\be
\boxed{
H_{[r^s]}^{{\rm db}_{m,n}}
=\det_{r\times r}R^{[r^s];(m,n)}
=\det_{s\times s}S^{[r^s];(m,n)}.
}
\label{eq:double-rectdet}
\ee
In matrix notation the same formula is
\be
H_{[r^s]}^{{\rm db}_{m,n}}
=\det_r\left(I_r+T^{(m)}M(T^{(n)})^tL\right)
=\det_s\left(I_s+M(T^{(n)})^tLT^{(m)}\right).
\ee
At $n=1$, (\ref{eq:Z-trefoil-factorization}) reduces the matrices
themselves to the twist matrices. The result is symmetric in $m,n$, although an
individual matrix entry in (\ref{eq:double-matrix-entries}) need not be.

\subsection{The first square example and the triple sum}

For $[2,2]$, use the four functions in (\ref{eq:T22-explicit}) and
\be
L=\operatorname{diag}\left(
-\frac{[2]^2D_2D_{-2}}{A^2},\,
\frac{[2]^2[3]D_2D_{-2}D_3D_{-1}}{A^4q^2}\right),\qquad
M=\operatorname{diag}\left(1,
-\frac{q^2[3]D_1D_{-3}}{A^2}\right).
\ee
Together these give an explicit $2\times2$ answer for all $m,n$.
For example, with $t_{ab}=T_{ab}^{(m)}$ and $u_{ab}=T_{ab}^{(n)}$,
\be
H_{[2,2]}^{{\rm db}_{m,n}}
=\det\begin{pmatrix}
1+\ell_1(t_{11}u_{11}+\mu_2t_{12}u_{12})&
\ell_2(t_{11}u_{21}+\mu_2t_{12}u_{22})\\
\ell_1(t_{21}u_{11}+\mu_2t_{22}u_{12})&
1+\ell_2(t_{21}u_{21}+\mu_2t_{22}u_{22})
\end{pmatrix}.
\ee

To derive the resummation, use
(\ref{eq:W-minors}), (\ref{eq:F-minor-ratio}) and
(\ref{eq:double-factorization}) to write the contribution of $\lambda$ as
\be
W_\lambda^{[r^s]}F_\lambda^{(m,n)}
=\frac{\det Z_{\mathsf A,\mathsf B}}
{\det T^{(1)}_{\mathsf A,\mathsf B}}
\det T^{(m)}_{\mathsf A,\mathsf B}
\det T^{(n)}_{\mathsf A,\mathsf B}.
\ee
The quotient of minors is especially simple:
\be
\frac{\det Z_{\mathsf A,\mathsf B}}
{\det T^{(1)}_{\mathsf A,\mathsf B}}
=\prod_{a\in\mathsf A}\ell_a\prod_{b\in\mathsf B}\mu_b.
\label{eq:reference-minor-cancellation}
\ee
All these denominators are nonzero at generic parameters, since
$T^{(1)}$ differs from the quantum Cauchy matrix $T$ only by row
and column factors. After their cancellation, Cauchy--Binet gives
(\ref{eq:double-rectdet}).

One must perform the cancellation at the level of minors.
Entry-wise multiplication and division instead give, in general,
\be
\det\left(\frac{T_{ab}^{(m)}T_{ab}^{(n)}}{T_{ab}^{(1)}}\right)
\ne\frac{\det T^{(m)}\det T^{(n)}}{\det T^{(1)}}.
\nn
\ee
The valid formula uses the two evolution matrices on opposite sides
of the contraction in (\ref{eq:double-matrix-entries}).

\subsection{Non-rectangular representations}

For a non-rectangular color, the additional channels must enter both
evolutions. In $[2,1]$, the first extra eigenvalue is the $A^4$
entry of (\ref{eq:R21-spectrum}); for $[3,2]$ there are twelve
reduced channels, including three off-diagonal composite pairs.
A full Racah matrix provides the corresponding double-evolution
polynomial through the arborescent construction. Extending the
determinant formula requires explicit matrices replacing
(\ref{eq:double-matrix-entries}) and a factorization replacing
(\ref{eq:Z-trefoil-factorization}); these remain to be constructed
for general non-rectangular colors.

\section{Conclusion}

The main result of this paper is that the large KNTZ evolution for
rectangular colors is the exterior power of a much smaller one-particle
system. In the notation of Appendix~\ref{app:matrix-origin},
\be
 \boxed{
 \Xi_R\mathcal B_R\Xi_R^{-1}
 =\omega_{\rm vac}^{-1}\operatorname{Comp}_s(g_R)
 =\omega_{\rm vac}^{-1}\wedge^s g_R, 
 }
 \label{eq:conclusion-exterior}
\ee
where the compound matrix ${\rm Comp}$ consists of all minors of $g_R$.
The one-particle matrix $g_R$ has size $(r+s)\times(r+s)$, whereas
$\mathcal B_R$ acts on the Young sub-diagrams of $[r^s]$ and has size
$\binom{r+s}{s}\times\binom{r+s}{s}$. Thus the shortest summary is
\be
 \boxed{
 \binom{r+s}{s}\times\binom{r+s}{s}
 \quad\longrightarrow\quad
 (r+s)\times(r+s).
 }
\ee

Appendix~\ref{app:matrix-origin} gives a direct finite-matrix explanation
of this reduction. The hook matrix $T^{(m)}$ is obtained from the
negative-to-positive block of $g^m$. A block multiplication of $g^m$
reduces its $m$-dependence to an ordinary geometric series, which explains
the factors $1-\lambda_{ij}^m$ in the explicit hook formula. The same
one-particle transfer can also be written as a diagonal factor times a
$q$-exponential of the elementary shift. In this form the complete and
skew Schur functions entering KNTZ arise directly from the coefficients
and minors of the shift operator.

For the figure-eight knot, Section~\ref{sec:41-cyclotomic} gives a
separate consequence of the determinant representation. At a primitive
root $q^2$ of order $\ell$, with $r,s<\ell$, the balanced matrix splits
beyond the wall $r+s=\ell$ into a residual rectangular block and
$r+s-\ell$ independent length-$\ell$ hook channels. The corresponding
factorization of $H^{4_1}_{[r^s]}$ is then an immediate determinant
identity. We do not claim the analogous block decomposition for general
twist and double-braid evolution here.

The operator form of $g$ also suggests a possible extension in which the
fixed specialization $p_n^\circ$ is replaced by independent shift times.
Whether the full knot observables then become KP/Toda $\tau$-functions is
a separate question.

\appendix
\section{The matrix origin of the evolution minors}\label{app:matrix-origin}

The determinant formulas in the main text use the explicit hook matrix
$T^{(m)}$.  This Appendix explains where this matrix comes from.  The
argument has three steps.  First, $T^{(m)}$ is obtained from powers of one
finite triangular matrix $g$.  Second, the factors $1-\lambda^m$ follow
from an ordinary geometric series for the blocks of $g^m$.  Finally, the
same matrix $g$ is identified with the usual KNTZ evolution by taking its
exterior power.  At the end we record a compact $q$-exponential form of
$g$, which makes the Schur structure transparent.

\subsection{The finite transfer matrix and the hook coefficients}

On the integer levels $-s,\ldots,r-1$, introduce
\be
\omega_k=A^{2k}q^{2k(k+1)},\qquad
g_{k\ell}=\begin{cases}
\displaystyle\omega_k\frac{(-1)^{k-\ell}
q^{-(k-\ell)(k-\ell-1)/2}}{[k-\ell]!},&k\geq\ell,\\[2mm]
0,&k<\ell.
\end{cases}
\label{eq:finite-transfer}
\ee
The matrix $g$ is lower triangular and has size $(r+s)\times(r+s)$.
Split the levels into the negative set $-s,\ldots,-1$ and the
nonnegative set $0,\ldots,r-1$.  If
\[
P^{(m)}=g^m,
\]
define the $r\times s$ matrix
\be
X^{(m)}=P^{(m)}_{+,-}\left(P^{(m)}_{-,-}\right)^{-1}.
\label{eq:X-graph}
\ee
The negative levels index the columns of $X^{(m)}$ in increasing order.
The hook matrix used in Section~\ref{sec:twists} is simply a diagonal
row--column rescaling of $X^{(m)}$:
\be
\boxed{
T_{ab}^{(m)}=(-1)^{b-1}q^{-c_{ab}}X^{(m)}_{a-1,-b}.
}
\label{eq:T-graph}
\ee
Thus all hook coefficients come from one evolved $s$-dimensional
subspace.  The columns of
$\left(\begin{smallmatrix}I_s\\X^{(m)}\end{smallmatrix}\right)$ form a
normalized frame of this subspace.  Replacing one negative level $-b$
by a nonnegative level $a-1$ gives a single hook coefficient; replacing
several levels gives the corresponding minor of $T^{(m)}$.

There is a simple reason for the dependence on the twist parameter $m$.
With respect to the same splitting of levels, write
\be
g=\begin{pmatrix}a&0\\ c&d\end{pmatrix}.
\label{eq:g-blocks-simple}
\ee
For $m>0$,
\[
(g^m)_{-,-}=a^m,
\qquad
(g^m)_{+,-}=\sum_{\nu=0}^{m-1}d^\nu c\,a^{m-1-\nu},
\]
and hence
\be
\boxed{
X^{(m)}=\sum_{\nu=0}^{m-1}d^\nu c\,a^{-\nu-1}.
}
\label{eq:X-geometric-series}
\ee
Both $a$ and $d$ are triangular.  Their eigenvalues are respectively
$\omega_{-j}$ and $\omega_{i-1}$.  Therefore every spectral component
of (\ref{eq:X-geometric-series}) contains the elementary geometric sum
\[
1+\frac{\omega_{i-1}}{\omega_{-j}}+\cdots+
\left(\frac{\omega_{i-1}}{\omega_{-j}}\right)^{m-1}.
\]
Since
\be
\frac{\omega_{i-1}}{\omega_{-j}}
=(Aq^{i-j})^{2(i+j-1)}=\lambda_{ij},
\label{eq:omega-ratio-lambda}
\ee
the numerator of this geometric sum is $1-\lambda_{ij}^m$.  This is the
origin of the factors appearing in the explicit formula
(\ref{eq:explicit-Tm}).  The remaining denominators and eigenvector
coefficients combine into the coefficients $C_{ab}^{ij}$ written there.
The same spectral formula extends to all integer $m$ because $g$ is
invertible at generic $A,q$.

\subsection{The exterior power and the KNTZ matrix}

All minors of the frame
$\left(\begin{smallmatrix}I_s\\X^{(m)}\end{smallmatrix}\right)$ are
therefore minors of a single evolved $s$-particle state.  This already
explains the minor ratio (\ref{eq:F-minor-ratio}).  To identify the same
finite construction with the customary triangular KNTZ matrix, set
\be
p_k^\circ=\frac{(q-q^{-1})^k}{q^k-q^{-k}},\qquad
\chi_\lambda^\circ=s_\lambda\{p^\circ\}
=\frac{q^{\sum_{\Box\in\lambda}c_\Box}}
{\prod_{\Box\in\lambda}[h_\Box]}.
\label{eq:pcirc-chi-simple}
\ee
The KNTZ matrix is \cite{1902.04140}
\be
{\cal B}_{\lambda\mu}
=A^{2|\lambda|}q^{4\sum_{\Box\in\lambda}c_\Box}
s_{\lambda/\mu}\{-p^\circ\}
\frac{\chi_\mu^\circ}{\chi_\lambda^\circ},\qquad
F_\lambda^{(m)}=({\cal B}^{m+1}{\bf1})_\lambda.
\label{eq:KNTZ-simple}
\ee
Here $-p^\circ$ means that every power sum changes sign.

For
\[
I_\lambda=\{\lambda_i-i:1\leq i\leq s\},
\]
Jacobi--Trudi writes the skew Schur factor in
(\ref{eq:KNTZ-simple}) as the corresponding minor of the triangular
part of $g$ after the occupied-level weights are removed.  The remaining
content factor telescopes:
\be
A^{2|\lambda|}q^{4\sum c_\Box}
=\frac{\prod_{k\in I_\lambda}\omega_k}
{\prod_{k=-s}^{-1}\omega_k}.
\label{eq:content-omega-simple}
\ee
Consequently conjugating ${\cal B}$ by
$\Xi_R=\operatorname{diag}(\chi_\lambda^\circ)$ gives the exterior power
of $g$ divided by the vacuum weight:
\be
\boxed{
\Xi_R{\cal B}_R\Xi_R^{-1}
=\omega_\emptyset^{-1}\wedge^s g
=\omega_\emptyset^{-1}\operatorname{Comp}_s(g),
}
\label{eq:KNTZ-exterior-power}
\ee
where $\omega_\emptyset=\prod_{k=-s}^{-1}\omega_k$.

There is one shift in the evolution parameter.  KNTZ uses $m+1$,
whereas (\ref{eq:T-graph}) uses $g^m$.  Put
$\Omega=\operatorname{diag}(\omega_k)$ and
$I_\emptyset=\{-s,\ldots,-1\}$.  The initial coordinates
$\chi_\lambda^\circ$ are the minors, with column set $I_\emptyset$, of
\be
(g^{-1}\Omega)_{k\ell}
=\frac{q^{(k-\ell)(k-\ell-1)/2}}{[k-\ell]!}
\quad(k\geq\ell).
\label{eq:g-inverse-components-simple}
\ee
After $m+1$ KNTZ steps this matrix becomes
$g^{m+1}g^{-1}\Omega=g^m\Omega$.  Therefore
\be
\chi_\lambda^\circ F_\lambda^{(m)}
=\omega_\emptyset^{-m}
\det(g^m)_{I_\lambda,I_\emptyset}.
\label{eq:evolved-occupation-minor}
\ee
The determinant of the negative block of $g^m$ is
$\omega_\emptyset^m$.  Dividing by this block gives exactly the
normalized frame in (\ref{eq:X-graph}).  Finally,
$\chi_\lambda^\circ=q^{\sum c_\Box}\det T_{\mathsf A,\mathsf B}$, and
the row and column factors in (\ref{eq:T-graph}) remove the same content
power.  This gives (\ref{eq:F-minor-ratio}).

\subsection{A compact operator form of the same transfer matrix}
\label{app:q-exponential-simple}

The previous derivation needs only finite matrices.  There is, however,
a useful compact way to write the same matrix $g$.  Let $|k\rangle$ be
the basis of the one-particle space and define
\[
N|k\rangle=k|k\rangle,\qquad
S|k\rangle=|k+1\rangle.
\]
The shift is truncated at the boundary, so $S^{r+s}=0$.  Put
\[
Q=q^2,\qquad
\Omega=A^{2N}q^{2N(N+1)},\qquad
(Q;Q)_n=\prod_{j=1}^{n}(1-Q^j),
\]
and define
\[
e_Q(z)=\sum_{n\geq0}\frac{z^n}{(Q;Q)_n}.
\]
Using
\[
[n]!=q^{-n(n-1)/2}\frac{(Q;Q)_n}{(1-Q)^n},
\]
the component formula (\ref{eq:finite-transfer}) becomes
\be
\boxed{
g=\Omega\,e_Q\bigl(-(1-Q)S\bigr).
}
\label{eq:g-q-exponential-simple}
\ee
Thus the non-diagonal part of $g$ is nothing but a $q$-exponential of
one elementary shift.

The same formula also explains the Schur functions in the previous
subsection.  Indeed,
\[
p_n^\circ=(-1)^{n-1}\frac{(1-Q)^n}{1-Q^n},
\]
so that, as a finite formal series in the nilpotent operator $S$,
\be
 e_Q\bigl(-(1-Q)S\bigr)
 =\exp\left(-\sum_{n\geq1}\frac{p_n^\circ}{n}S^n\right)
 =\sum_{d\geq0}h_d\{-p^\circ\}S^d.
\label{eq:qexp-schur-simple}
\ee
Hence
\[
g_{k\ell}=\omega_k h_{k-\ell}\{-p^\circ\},
\]
and Jacobi--Trudi turns its minors into the skew Schur functions
$s_{\lambda/\mu}\{-p^\circ\}$ appearing in KNTZ.  In other words, the
$q$-exponential form is not an additional construction: it is a compact
operator rewriting of the same finite transfer matrix used above.

At a root of unity the coefficients of this $q$-exponential themselves
become singular when $(Q;Q)_\ell=0$.  We therefore do not use
(\ref{eq:g-q-exponential-simple}) to take the cyclotomic limit.  The
regular factorization of the figure-eight polynomial is instead obtained
from the balanced matrix in Section~\ref{sec:41-cyclotomic}.

\end{document}